\documentclass[12pt]{article}
\usepackage{amssymb,amsmath,epsfig}
\begin{document}

\begin{center}
\Large\textbf {Thermodynamic analysis of magnetic phase transitions in the ferromagnetic superconductor UGe$_{2}$ at ambient pressure}

\medskip

\normalsize
\textbf{Diana V. Shopova}
\end{center}
\medskip

\begin{center}
\normalsize
Institute of Solid State Physics, Bulgarian Academy of Sciences, \\
1784 Sofia, Bulgaria, sho@issp.bas.bg
\end{center}

 \medskip

\begin{abstract}
We study the possibility to apply phenomenological approach to the description of magnetic transitions in UGe$_2$ at ambient pressure with the help of Landau free energy expanded to $8^{th}$ order in magnetisation. The analysis shows that for certain values of parameters in front of $M^4$, and $M^6$ terms in the free energy there is possibility for the appearance of two successive phase transitions between the low-magnetisation and high-magnetisation phases with the same magnetic structure. We establish the relation of  the parameters in Landau energy, for which the phase transition from the disordered to low-magnetisation phase is of second order as the experimental data shows. We have also found the relations between the Landau coefficients which result in transition of crossover type between the low- and high-magnetisation phases when temperature is lowered.
\end{abstract}



\section{\label{1} Introduction}

\begin{figure}[!ht]
\begin{center}
\includegraphics[scale=0.55]{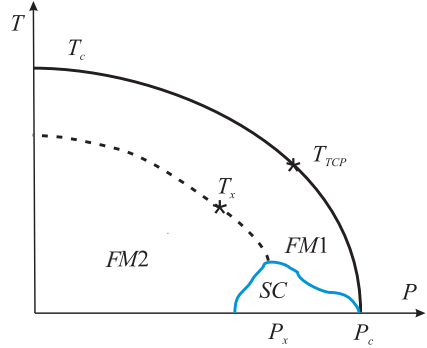}
\end{center}
\caption{\small{The schematic temperature-pressure phase diagram of UGe$_2$. FM1 and FM2 describe the low- and high-magnetisation phases, respectively.  The blue curve and the area below it describe the superconducting phase.}}
\label{Fig1}
\end{figure}

It is established that in Uranium-based ferromagnetic superconductors UGe$_2$, URhGe, and UCoGe~\cite{Pfleiderer:2009},~\cite{Saxena:2000}, ~\cite{Gasparini:2007}, the same  $5f$-electrons are responsible for  the occurrence of both ferromagnetism and superconductivity. The first discovered ferromagnetic superconductor is UGe$_2$, where the superconductivity appears only under pressure deep inside the ferromagnetic phase at low temperature. There exist two different ferromagnetic phases with same magnetic structure and different magnetisations. The crystal structure of UGe$_2$ is orthorhombic  where the U-atoms form coupled zigzag chains along the a-axis and the magnetic moments are aligned along the direction of the chain. In~(\ref{Fig1}) the schematic pressure-temperature phase diagram for UGe$_2$ is shown, and recent summarized experimental diagram may be found in ~\cite{Tateiwa:2018}. The transition at ambient pressure from paramagnetic to weakly polarized ferromagnetic phase (FM1)occurs at T$_c$ = 52.6 K  through second order phase transition with M$_0 = 0.9 \mu_B$/U. With the increase of pressure, the transition changes from second to first order at tricritical point T$_{CP} \simeq 22 $ K,and  $P_{CP} \approx 1.42$ GPa and ferromagnetism and superconductivity  disappear at pressure $\sim 1.5$ GPa and T=0.\\
When  temperature is lowered a second ferromagnetic phase (FM2) appears with the same magnetic structure and stronger magnetic polarization -1.41$\mu_B/$U ~\cite{Tateiwa:2014} at T$_x\approx$ 30 K. (see~(\ref{Fig1})). It is experimentally found that at ambient  pressure the transition between FM1 and FM2 ferromagnetic phases is of crossover type and only with increasing pressure there is a true first order phase transition close to the appearance of superconductivity with a maximum transition temperature of $\approx 0.7$ K at p$_x$. The superconductivity exists in a limited pressure range between 1.0 and 1.5 GPa. In this pressure range, the magnetic  moment is $1 \mu_B/U$ and the transition between two ferromagnetic phases is of first order at the critical end point (CEP: T$_{CEP} \sim 7 $ K, P$_{CEP} \sim 1.16 $ GPa).\\
Experimentally, it is established that UGe$_2$ has very strong uniaxial magnetic anisotropy~\cite{Onuki:1992}, and at some earlier stages of study, it has been considered that 3D Ising model can well describe its magnetic properties - see, for example,~\cite{Shopova:2009}, and for recent review ~\cite{Aoki:2019}. It is experimentally found and theoretically  justified~\cite{Tateiwa:2014} that the  critical exponents for UGe$_2$ do not belong to any known universality class including 3D Ising model and anisotropic next nearest neighbor 3D Ising model. One of the reasons for this lays in the dualism of 5f-electrons in UGe$_2$ where two subsets of itinerant and localised ones exist. This fact is experimentally established by positive muon relaxation measurements where the  itinerant magnetic moment at ambient pressure is found to be $\approx $ 0.02$ \mu_B$ ~\cite{Yaouanc:2002}. Later this dualism of 5f-electrons is confirmed by magnetic, transport and specific heat experiments, see, for example, ~\cite{Troc:2012}. \\ Understanding of magnetic phase transitions from microscopic point of view is important for understanding the appearance of superconductivity - for review, see for example ~\cite{Huxley:2015}, but there is now up to now consensus on this problem. Recently, an estimate is made on the basis of Takahashi's spin fluctuation theory about the itinerancy of 5f-electron systems in actinides~\cite{Tateiwa:2017}. The comparison of this phenomenological approach with the experimental data shows that UGe$_2$ within the criterion used in the paper is an intermediate case between strongly localised and itinerant 5f-electron system. In the paper~\cite{Tateiwa:2018}, measurements of magnetization are performed at high pressure for UGe$_2$ with the aim to make clear the correlation between the superconductivity and pressure-enhanced ferromagnetic fluctuations. From the experimental data a conclusion is made that at lower pressure the ferromagnetic state is more itinerant in FM1 compared to FM2. \\
Usually microscopic models are focused on the phase transition between FM1 and FM2 around the critical end point, where superconducting phase appears under pressure ~\cite{Fidrysiak:2019} although at ambient pressure this problem is not yet resolved, especially
of the nature of crossover transition between FM1 and FM2. In earlier studies, it is proposed that the electronic density of states has two very closely placed peaks near the low-dimensional anisotropic Fermi surface, whose topology  changes by the appearance of magnetisation ~\cite{Sandeman:2002}. This calculation is performed at T=0 with the help of Stoner theory for itinerant magnets. There the authors claim than within mean-field approximation their theory will be in analogy with Shimizu paper~\cite{Shimizu:1982} for the metamagnetic transitions under external magnetic field by including the  $M^8$ term in the free energy. The experimentally established first order FM1$\rightarrow$FM2 around $p_x$ is also considered as metamagnetic, but in UGe$_2$ this metamagnetism exhibits some peculiar features, for example T$_x$ $\ll$T$_c$ and no reorientation or reversal of spin occurs, but only change in the magnitude of magnetic moment. \\\\
In this paper we will consider the ferromagnetic phase transitions in UGe$_2$ at ambient pressure in the absence of external magnetic field. There are few experimental data in this  regime, see for example  the experiments on linear thermal expansions of single-crystalline UGe$_2$,~\cite{Oomi:2002} at ambient pressure. Usually the transition between FM1 and FM2 at pressures lower than 1 GPa down to $p=0$ is considered as crossover, see, for example~\cite{Tateiwa:2018} and~\cite{Aoki:2019} and not a real thermodynamic phase transition. At ambient pressure there are experimentally observed singularities at $T_x=30$ K in different physical quantities, for example broad hump in the specific heat, see \cite{Pfleiderer:2009} and the papers cited therein. But systematic measurement at ambient pressure which may shed light on the nature of FM1$\rightarrow$FM2 crossover at $p=0$ are missing. Special attention can be paid to positive muon spin rotation measurements of paper~\cite{Sakarya:2010} where the authors claim  that at pressures 1.00 GPa and below the FM1$\rightarrow$FM2 corresponds to a real thermodynamic phase transition, but there the explanation of obtained experimental data at low and especially at ambient pressure is not quite definite.  \\
In view of the above arguments we will adopt here the hypothesis that at ambient pressure and zero magnetic field the transition between two ferromagnetic phases in UGe$_2$ is of isostructural nature as defined in~\cite{Gufan:1978} and later on generalised see, for example \cite{Izyumov:1984}, \cite{Toledano:1987}. This means that the phase transition may occur between two ordered phases with the jump of order parameter without changing the structure (for example, the change of volume and/or valence without change of crystal symmetry). From the general theory for isostructural phase transitions, they can be described by including in the expansion of Landau free energy, a term to the 8$^{th}$ order in magnetization, namely, M$^8$. For this free energy expansion the phase diagram is rich and proposes different solutions depending on the values of coefficients in the Landau expansion. Here we find out those relations between the coefficients which can lead to the experimentally observed crossover between FM1 and FM2 in UGe$_2$, and the phase transition from paramagnetic to FM1 phase is of second order. Such an approach gives further possibility to include in calculations the influence of external pressure by supposing the model dependence of Landau coefficients on it.\\

 \section{\label{1} Theoretical approach}

\begin{figure}[!ht]
\begin{center}
\includegraphics[scale=0.45]{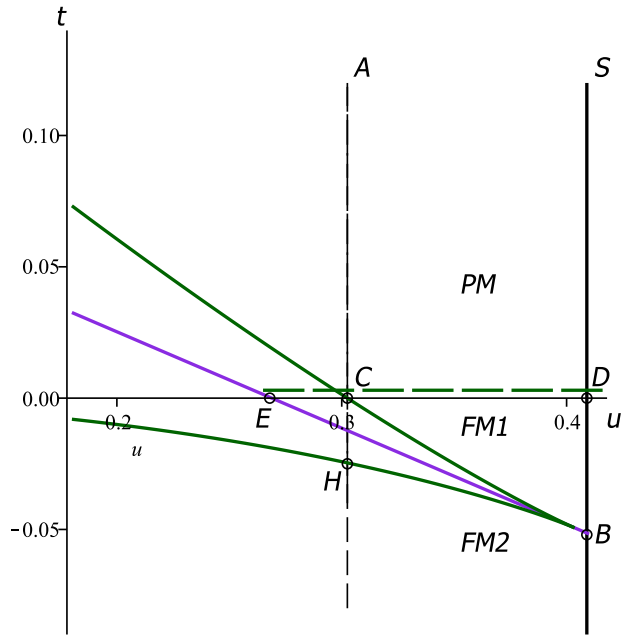}
\end{center}
\caption{\small{The solutions $t_{1,2}$ of quantity Q as function of the parameter $u$. The dashed vertical curve is drawn at $u=\frac{w^2}{4}$ marked by point C in the figure. Point B is at $u=\frac{w^2}{3}$.}}
\label{Fig2}
\end{figure}

As pointed in the previous section the magnetization in UGe$_2$ is highly anisotropic due to localized 5f-electrons. The a-axis is the easy magnetization axis along which  $M_a$, coming from both itinerant and localised 5f-electrons has the greatest value. The magnetic moment coming from itinerant 5f-electrons is isotropic and, in principle, also the perpendicular part of magnetisation ($M_b$, $M_c$), where b and c are the crystal axes, should be taken into account. \\
As a first step in considering the phase transitions between the paramagnetic phase and FM1 and FM2 at ambient pressure we will not take into account  the transverse magnetisation as small compared to the magnitude of magnetisation along  $a$-axis - $\sim 0.02$ as experiment shows, and will expand the Landau free energy $F(M)$ in $\overrightarrow{M}=(0,0,M_a)$, the subscript $a$ will be omitted hereunder.
Then $F(M)=f(M)V$ where $V$ is the  volume and the free energy density $g(M)$ is given by the expression:
\begin{equation}\label{Eq1}
g=a M^2 +\frac{b}{2}M^4 +\frac{c}{3}M^6+\frac{v}{4}M^8.
\end{equation}
In the above equation, $a=\alpha (T-T_c)$, where $T_c$ is the Curie temperature and $\alpha$ is a material parameter. The other coefficients in the Landau expansion, namely, $b, \; c, \; v$ are considered at this stage not dependant on temperature and external pressure, namely $b=b(T_c,P_0),\;c=c(T_c,P_0), \;v=v(T_c,P_0)$; here by $P_0$ we denote the ambient pressure. The sign of $b, \; c$ may be either positive or negative, but $v>0$ in order to ensure convergence of $f$.\\
In the expression (\ref{Eq1}) there are too many unknown parameters, which in principle may be derived from experiment in combination with respective microscopic calculations at ambient pressure. In order to reduce this number, as well as to make free energy density $f$ dimensionless  we will redefine  the order parameter $M$ by introducing:
\begin{equation}\label{Eq2}
m=v^{1/8}M.
\end{equation}
Then the  free energy density - Eq. (\ref{Eq1}) in dimensionless form becomes:
\begin{equation}\label{Eq3}
  f= tm^2+\frac{u}{4}m^4+\frac{w}{3}m^6+\frac{1}{4}m^8;
\end{equation}
and the coefficients in Eq.(\ref{Eq3}) are related to the initial ones in the following way:
\begin{eqnarray*}
\nonumber u &=& \frac{b}{v^{1/2}}\\
\nonumber  w &=& \frac{c}{v^{3/4}}
\end{eqnarray*}
The reduced temperature is $t =\beta (T/T_c-1)$ with $T_c$ - the Curie temperature, $\beta=\alpha T_c/v^{1/4}$.\\
The equation of state $(df/dm)$ :
\begin{equation}\label{Eq4}
2m(t+um^2+wm^4+m^6)=0
\end{equation}
has an obvious solution for disordered (paramagnetic) phase $m=0$.\\
The stability condition is given by inequality:
\begin{equation}\label{Eq5}
\frac{d^2f}{dm^2}=2(t+3um^2+5wm^4+7m^6)\geq 0
\end{equation}
and there are several methods to resolve this condition in analytical form: see for example ~\cite{Izyumov:1984}. This may be also done numerically by direct substitution of solutions of Eq. (\ref{Eq4}) in the above equation and analysing its positiveness. If more than one solution is stable also a comparison between free energies of respective solutions of Eq. (\ref{Eq4}) should be made in order to find out which one in what domain of reduced temperature is an absolute minimum. \\
If we substitute $x=m^2 \geq 0$, the equation (\ref {Eq4}) will become standard 3-rd order algebraic equation, its solutions can be given in analytical form, see, for example~\cite{Abramowitz:1964}.\\ Eq. (\ref{Eq4}) expressed by new variable $x$ reads:
\begin{equation}\label{Eq6}
t+ux+wx^2+x^3=0.
\end{equation}
The number of real solutions for $x=m^2$ is determined by the sign the quantity
\begin{equation}\label{Eq7}
Q=\frac{t^2}{4} +\frac{2w}{3}(\frac{2w^2}{9}-u)t+\frac{u^2}{27}(u-\frac{w^2}{4})
\end{equation}
The quantity $Q(t)$ which is a quadratic equation with respect to the reduced temperature, $t$ as function of parameters $u,w$ with solutions $t_{1,2}$ given by:
\begin{equation}\label{Eq8}
t_{1,2}=\frac{w}{3}(u-\frac{2}{9}w^2)\pm \frac{2}{27}(w^2-3u)^{3/2}
\end{equation}
also determines the region of existence of non-negative solutions for magnetisation as function of parameters $u$ and $w$ in Landau energy. It is obvious that $t_{1,2}$ are real for $w^2\geq 3u$ and $t_0=w^2/(3u)$ is a special point, for which $t_1=t_2=w^3/27$.
Q may also be written as
\begin{equation}\label{Eq9}
Q=(t-t_{1})(t-t_{2}).
\end{equation}
Independent of the sign of $w$, $t_1>t_2$; then for $t$ between t$_{1,2}$, Q$<0$ and Eq. (\ref{Eq6}) will have three real solutions; for $t> t_{1}$, and $t< t_{2}$, Q$>0$ and there are one real solution and two complex conjugate solutions; for Q$=0$ - three real solutions, two of them are equal. In addition, the real solutions should be non-negative as $x=m^2$ and this strongly depends both on sign and magnitude of parameters $w$ and $u$.\\
We denote the solutions of Eq. (\ref{Eq6}) by $x_1,\; x_2, \; x_3$.\\
The description of all solutions of equation of state, Eq. (\ref{Eq4}), is very well illustrated in the parameter space $(u,t)$ for fixed value of coefficient $w$. We show graphically in Fig. (\ref{Fig2}) the dependence of $t_{1,2}$ on parameter $u$, for fixed parameter $w<0$. In this case $t_2<0$; $t_1$ is positive  and intersects the $u$-axis at $u=w^2/4$.\\
The $(u,t)$ diagram for $w>0$ is mirror image of Fig. (\ref{Fig2}) as there $t_1(w<0)=-t_2(w>0)$, and $t_2(w<0)=-t_1(w>0)$; the point B is above $u$-axis, and $t>0$.\\ We will not consider positive values of parameter $w$ as the analysis shows that for $w>0$, and $u>0$, no isostructural transition appears and only one ferromagnetic phase exists through second order transition from the para phase. In this case to expand Landau free energy up to $M^8$ is redundant. For $w>0$ and $u<0$ the ferromagnetic phase occurs through first order phase transition from para phase and again no isostructural transition may occur. \\ The experiment shows that in UGe$_2$ the phase transition para$\rightarrow$ FM1 is of second order at ambient and low pressures. The order of phase transition to FM1 changes from second to first only under increasing pressure, passing through tricritical point (see Fig. (\ref{Fig1}), point (T$_{TCP}$)).\\
Here we will make some remarks on the appearance of tricritical point T$_{TCP}$ on the transition line between disordered phase and FM1 with the increase of pressure from zero to $P_c$, at which the ferromagnetism disappears. Within the Landau approach we use this means change of parameter $u$ with pressure from positive to negative value. At the tricritical point this parameter turns zero, see \cite{Toledano:1987}. In order to describe the change of order of transition para$\rightarrow$FM1 with pressure, a supposition should be made on the dependence of $u$ on pressure based on microscopic picture of behavior of f-electrons with pressure.  \\Another theory shows that the effects of gapless particle-hole excitations at the Fermi surface for itinerant ferromagnets induce a nonanalytic term of logarithmic type in the Landau expansion of the free energy as a function of the magnetic moment M  \cite{Belitz:1999}. When applied to magnetic transitions in UGe$_2$ \cite{Belitz:2005} this theory explains the appearance of tricritical point and the first order transition at low temperature, as well as the second order transition  at higher temperatures. In a recent review on the U-based ferromagnetic superconductors \cite{Aoki:2019} different theoretical approaches to the UGe$_2$ phase diagram under pressure are presented. The discussion of dependence of magnetic phase transitions in UGe$_2$ on pressure is outside the scope of our paper.\\
Further we will consider only $w<0$, for which two ferromagnetic phases may appear with the decrease of temperature.
The calculations show that the solution $x_1$ of equation of state (\ref{Eq6}) is always positive and bigger than $x_2$; $x_2$ may be negative for $t>0$ and positive for $t<0$. This is why, for $w<0$ we consider that $x_1$ describes the high-magnetization phase and $x_2$ - the low-magnetization phase. The solution $x_3$ is positive and unstable.  When $u\leq 0$ and $w<0$, $x_1$ exists and is positive even for $t>0$,  $x_2$ is negative, $x_3$ is positive but metastable in the region between $t_1$ and $t_2$ according to the stability condition, Eq. (\ref{Eq5}) and no isostructural transition can take place there. The phase transition from paramagnetic to ferromagnetic phase is of first order.\\
The calculations show that region of interest  in the $(u,t)$ phase diagram, see Fig.~(\ref{Fig2}) is where $Q\leq 0$, for negative values of coefficient $w<0$ before $m^6$, see Eq.(\ref{Eq3}) and positive values of coefficient $u$ before $m^4$ as far as there is possibility for the occurrence of transition between two magnetic phases with same structure.\\
We will describe in some detail different points and lines in  Fig.~(\ref{Fig2}).
The violet line in the figure denotes the equilibrium line $t_{eq}$ of first order isostructural phase transition and  is determined by making equal the free energies $f$, Eq. (\ref{Eq3}), of low- and high-magnetisation phases $f(m_1)=f(m_2)$.
The equilibrium line $t_{eq}=\frac{w}{3}\left(u-\frac{2}{9}w^2\right)$ crosses the $u$-axis at $u_E=\frac{2}{9}w^2$, denoted by point E in Fig (\ref{Fig4}). For $0<u<\frac{2}{9}w^2$, the high-magnetization phase ($x_1$) is more stable than the low-magnetization phase ($x_2$) in the whole interval of its existence and occurs through first order phase transition from para phase; this transition is shifted to positive values of $t$. In this interval of $u$ and within our approximation the low-magnetization phase always has higher free energy than the high-magnetization phase and is metastable.\\
For $u=u_E$ the high-magnetization phase is stable only for $t<0$ but $f(x_2)$ is higher than $f(x_1)$ and no isostructural transition can occur within our approximation. When $u>u_E$, the region of stability of $x_1$ is shifted to more negative values of $t$ and in the interval of $0<t \leq t_{eq}$, $x_2$ has lower energy than $x_1$ and this interval grows with the increase of $u$.\\
Another special point in Fig. (\ref{Fig4}) is at $u=w^2/4$, point C where $t_1$ crosses $u$-axis. There $x_1$ is non-negative only for $t\leq 0$, and its stability is shifted to $t<0$. As far as the FM1 in UGe$_2$ exists in a relatively large temperature interval $\sim 22$ K, we are interested in those values of $u$ where the low magnetization phase, described by $x_2$ exists and is stable in relatively large interval of reduced temperature $t$. The calculations show that this condition is fulfilled for $w^2/4<u<w^2/3$.
In this range of parameter $u$ the phase transition from the disordered to ferromagnetic phase is of second order and when the temperature is lowered isostructural transition  of first order occurs.\\
The  region of parameters $u$ where second order phase transition from disordered phase followed by first order isostructural transition to high-magnetization phase can occur is illustrated in Fig. (\ref{Fig4}) by green dashed line.\\

\section{\label{3} Results and discussion}

\begin{figure}[!ht]
\begin{center}
\includegraphics[scale=0.5]{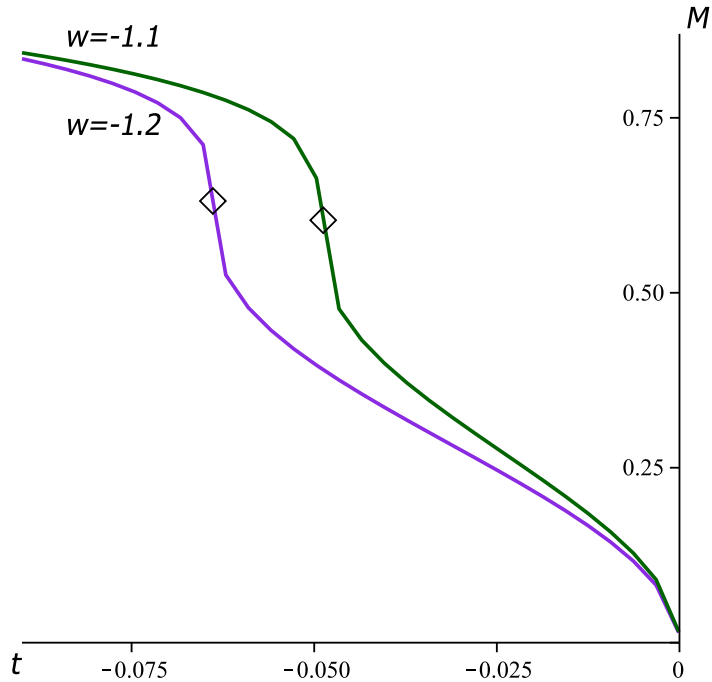}
\end{center}
\caption{\small{The dependence of magnetisation $M$ on the reduced temperature $t$ for two values of the parameter $w$. The points marked by diamonds on the curves denote the  crossover transition between FM1 and FM2.} }
\label{Fig3}
\end{figure}

\begin{figure}[!ht]
\begin{center}
\includegraphics[scale=0.5]{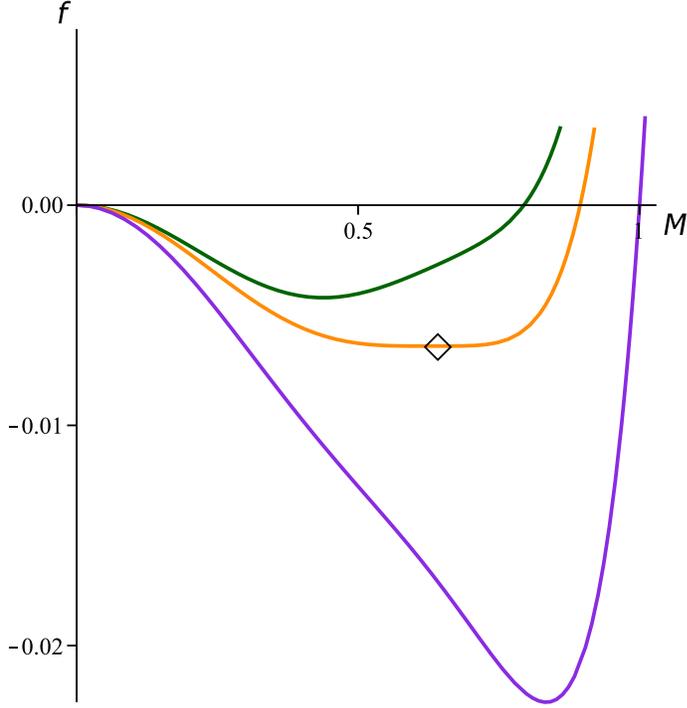}
\end{center}
\caption{\small{The dependence of free energy $f_c$ , Eq.(\ref{Eq10}) on magnetisation $M$ at different values of reduced temperature $t$. The curves are drawn for $w=-1.2$.The orange curve is for $t_{cr} = \frac{w^3}{27}$; the point denoted by diamond marks the value of $f$ for  $M_{cr}=\sqrt{-3w}/3$. The green curve is drawn for $0>t>t_{cr}$, and the violet line is for  $t<t_{cr}$. }}
\label{Fig4}
\end{figure}

\begin{figure}[!ht]
\begin{center}
\includegraphics[scale=0.4]{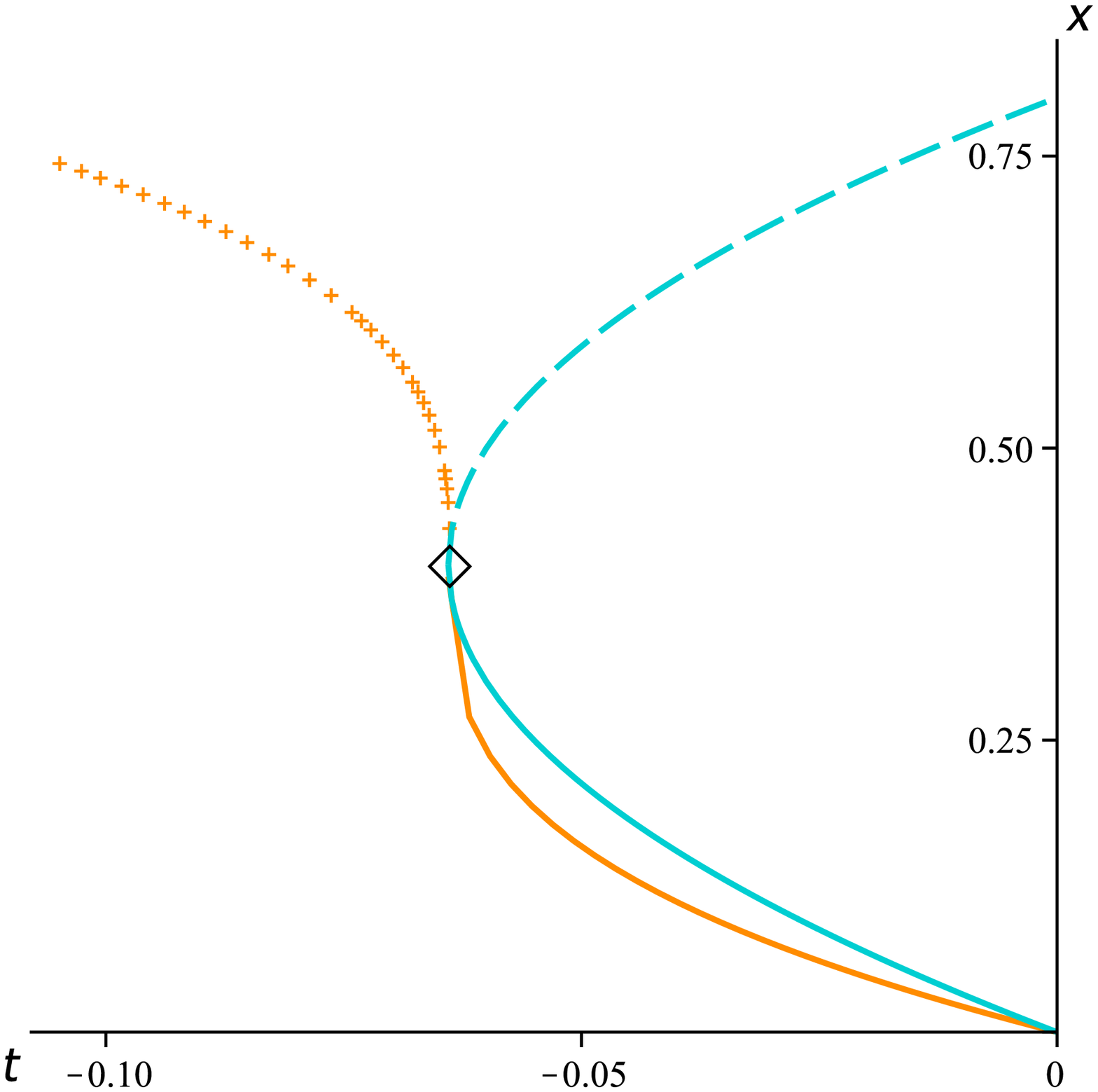}
\end{center}
\caption{\small{The stability conditions for FM1 and FM2. On the vertical axes $x=M^2$ and the blue lines denote $M^2_{1,2}$ as given by Eq.(\ref{Eq12}). The solid orange line stands for $M^2$ of low-magnetisation phase, and the orange curve marked by crosses is for $M^2$ of high-magnetisation phase. The diamond marks $M^2$ at $t_{cr}$.  }}
\label{Fig5}
\end{figure}

\begin{figure}[!ht]
\begin{center}
\includegraphics[scale=0.4]{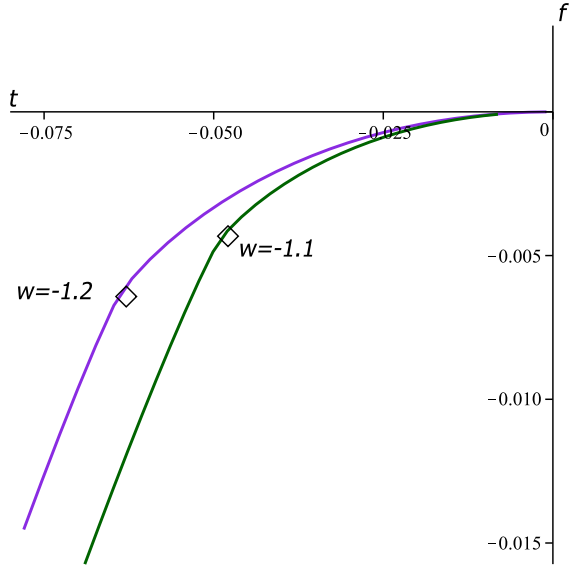}
\end{center}
\caption{\small{The dependence of equilibrium free energy $f_c$ , Eq. (\ref{Eq10}) on reduced temperature $t$. The points marked by diamonds show the crossover transformation between the low- and high-magnetisation phases. }}
\label{Fig6}
\end{figure}

The possibility for second order phase transition to low-magnetisation phase followed by isostructural transition to high magnetisation phase includes the values of $u$ at fixed $w$ in the interval $w^2/4<u\leq w^2/3$; points $C$ and $D$ in Fig.(\ref{Fig2}), respectively. The calculations show that within this interval with exception of point $D$ there is real first order transition between  FM1 and FM2 and this contradicts to the experiments in UGe$_2$ at ambient pressure, which claim that there is not a true thermodynamic transition FM1-FM2 at ambient pressure, but a crossover, see for example the review \cite{Aoki:2019}. So we suppose that at ambient pressure the Landau expansion to $M^8$ can be applied to description of FM1-FM2 transition if we adopt that within our approach that the following conditions are fulfilled $u>0$, $w<0$ and $u=w^2/3$ for the Landau coefficients in the free energy expansion Eq.(\ref{Eq2}). This is illustrated in Fig.(\ref{Fig2}) by the line $SDB$. At $t=t_D=0$, the transition of second order from para phase to low-magnetisation phase occurs. At $t=t_B=w^3/27$ there is a crossover between FM1 and high-magnetisation phase. FM1 exists and is stable in the interval $BD$ and for $t<t_B$, the high-magnetisation phase is stable.\\
To be more explicit we will write down the expression of free energy for $u=w^2/3$, denoted by $f_c$ and the respective equation of state:
\begin{equation}\label{Eq10}
f_c=tM^2+\frac{1}{6}w^2 M^4+\frac{1}{3}w M^6+\frac{1}{4}M^8,
\end{equation}
and\\
\begin{equation}\label{Eq11}
\frac{\partial f_c}{\partial M}=\frac{2}{3} M(3t+M^2 w^2+3M^4w+3M^6)=0
\end{equation}
For $0<t<w^3/27$, there is one real positive solution $M_1$ for magnetisation of Eq. (\ref{Eq11}) and  in the temperature interval $t<w^3/27$ , the real positive solution for magnetisation is $M_2$ ; the calculations show that $M_1<M_2$. At the crossover reduced temperature, denoted by $t_{cr}=w^3/27$, the equation of state, Eq. (\ref{Eq11}), becomes:
$$\frac{1}{9}(3M^2+w)^3$$ with the obvious solution for magnetisation $M^2=-w/3$ and all real positive solutions for $M$ are equal.
The dependence of magnetisation on the reduced temperature is illustrated in Fig. (\ref{Fig3}).\\
It is seen from the figure that around $t_{cr}=w^3/27$ there is a jump in magnetisation of FM2 phase. This is reflected on the behaviour of free energy as function of magnetisation for different values of reduced temperature $t$, as shown in Fig. (\ref{Fig4}).\\
There is a very broad and flat  minimum at $t_{cr}$, see the orange curve in Fig.(\ref{Fig4}), where the diamond denotes the equilibrium magnetisation as a solution of equation of state Eq. (\ref{Eq11}) at crossover transition. The minimum of $f_c$ for $0>t>t_{cr}$ gives the equilibrium value of $M$ for low-magnetisation phase, and the respective minimum for $t<t_{cr}$ the equilibrium value of $M$ for high-magnetisation phase  \\
We should mention here that the second derivative of free energy  Eq. (\ref{Eq10}) defines the stability of ferromagnetic phases. Taking into account the equation of state, Eq. (\ref{Eq10}), the stability condition may be written in the form
$$-4w(M^2-M_1^2)(M^2-M_2^2)\geq0$$ with
\begin{equation}\label{Eq12}
M^2_{1,2}= \frac{-w}{3} (1\pm\sqrt{1-\frac{27}{w^3}t})
\end{equation}
and it is satisfied when $M^2>M_1^2$ or $M^2<M_2^2$.\\ The calculations show that the high-magnetisation phase satisfies the first condition and the low-magnetisation phase - the second one within the respective temperature intervals. For $M_{cr}$ the stability condition is equal to zero. The stability conditions are illustrated graphically in Fig. (\ref{Fig5}) for $w=-1.2$. The area encircled by blue lines is the region of instability. At temperature of crossover transition, the stability is marginal, as both second and third derivatives of free energy Eq.(\ref{Eq10}) are zero there. \\
This is seen very well in the dependence of equilibrium free energy Eq. (\ref{Eq10}) on the reduced temperature $t$ shown in Fig. (\ref{Fig6})for two values of $w$.\\
There is a small cusp at $t_{cr}$ on the curves, which marks the crossover between FM1 and FM2.\\
We suppose that taking into account also the transverse components of magnetisation the above established relation between the coefficients $u$ and $w$ in Landau expansion, namely $u=w^2/3$ will be modified. From experimental data the ratio between transverse and longitudinal magnetisations is $\sim 0.02$, so the transverse components may be included as perturbation and such approach is justified. Although small, their role in description of ferromagnetism and especially of superconductivity is important as pointed, for example in~\cite{Pfleiderer:2009}.

\section{\label{4}  Concluding remarks}
In this paper we  apply the phenomenological Landau approach to UGe$_2$ for the description of magnetic phase transitions at ambient pressure far from superconducting transition. The aim is to find a model, by which the occurrence of two different magnetic phases with same magnetic structure but different magnetisations can be described. We propose expansion of Landau free energy in magnetisation up to $M^8$, for which by general theory it has been proven that for particular values of Landau parameters, there is a possibility for isostructural transition to occur,~\cite{Gufan:1978}.~\cite{Izyumov:1984},~\cite{Toledano:1987}. We found the relations between the parameters of Landau energy for which there is possibility of second order phase transition to low-magnetization phase with subsequent transition of crossover type at temperature lowering to high-magnetization phase with the same magnetic structure. As shown there is strong dependence between the magnitudes of parameters $u$ and $w$, see Eq. (\ref{Eq6}), and obviously for the appearance of isostructural transition from low- to high-magnetisation phase $ w<0,u>0$ and $u=w^2/3$ . In order to make reliable estimate of the magnitude of these parameters an adequate microscopic calculations should be made for magnetic phase transitions at ambient pressure.
\\
Also the fact of working in quite rough approximation, dropping the temperature dependence of coefficients $u$ and $w$ before $m^4$ and  $m^6$ terms in Landau energy, Eq. (\ref{Eq3}) limits the possible variations of both $w$ and $u$. \\The nature of magnetic phase transitions from experimental point of view at $p=0$ is still an open question. Here we propose  a general thermodynamic approach which does not depend on the undelying microscopic theories  but we consider that at this first stage of our study to make direct comparison with experiment is preliminary. Our model needs some generalisation, especially the coefficient in the Landau free energy, Eq. (\ref{Eq1}), namely, $b=b(T_c,P_0)$; should be considered as dependent on temperature and not fixed at Curie temperature.\\
As far as it is generally accepted that the spin fluctuations are responsible for occurrence of superconductivity of p-type, it will be of great interest also to consider their role on magnetic transitions within this phenomenological approach. Recent experiments at ambient pressure \cite{Noma:2018} show that the spin fluctuations in UGe$_2$ are highly anisotropic, and the authors claim that in PM state longitudinal spin fluctuations prevail, while in FM state the transverse spin fluctuations are predominant. The nature of spin fluctuations at ambient pressure remains unclear, having in mind also the dual nature of f-electrons. The above issues will be the subject of future studies.

\section*{Acknowledgements}
This work is supported by Grant $KP-06-N38/6$ of the Bulgarian National Science Fund.

\end{document}